\documentclass[a4paper,10pt,oneside]{article}
\usepackage[latin1]{inputenc}
\usepackage{amsmath}
\usepackage{amsfonts}
\usepackage{amssymb}
\usepackage{graphicx}
\usepackage{footnote}
\usepackage{float}
\usepackage{subcaption}
\DeclareGraphicsExtensions{.bmp,.png,.pdf,.jpg,.eps}
\usepackage{physics}
\usepackage{mathrsfs}
\usepackage{cite}
\usepackage[colorlinks,urlcolor=blue,citecolor=blue]{hyperref}

\advance \textheight by \topskip
\usepackage{amsmath}
\usepackage[english]{babel}
\makeatletter
 \@addtoreset{equation}{section}
\makeatother
\usepackage[latin1]{inputenc}
\usepackage{amsmath}
\numberwithin{equation}{section}

\advance \textheight by \topskip
\usepackage{amsmath}
\usepackage[english]{babel}
\makeatletter
 \@addtoreset{equation}{section}
\makeatother
\usepackage[latin1]{inputenc}
\usepackage{amsmath}
\numberwithin{equation}{section}

\advance \textheight by \topskip
\usepackage{amsmath}
\usepackage[english]{babel}
\DeclareMathAlphabet\mathbfcal{OMS}{cmsy}{b}{n}
\DeclareMathAlphabet{\boldmathe}{T1}{cmr}{bx}{it}
\newcommand{\mbf}[1]{\boldmathe{#1}}
\newcommand{\mbfgr}[1]{\textit{\mbox{\boldmath$#1$}}}

\def\vK{\mbf{K}}

\def\vj{\mbf{j}}

\def\vr{\mbf{r}}

\def\vsigma{\,\mbfgr{\sigma}}

\def\be{\begin{equation}}
\def\ee{\end{equation}}

\def\R{\mathbb R}

\def\be{\begin{equation}}
\def\ee{\end{equation}}

\def\R{\mathbb R}

\def\vK{\mbf{K}}

\def\vr{\mbf{r}}

\def\vsigma{\,\mbfgr{\sigma}}
\def\vvarphi{\,\mbfgr{\varphi}}

\def\be{\begin{equation}}
\def\ee{\end{equation}}

\def\R{\mathbb R}

\def\Ti{\text{i}}

\usepackage{todonotes}

\begin{document}

\begin{center}
{\LARGE \bf
	SUSY design of smooth quantum rings in graphene\\ 
}
\vspace{6mm}
{\Large Francisco Correa$^a$, Luis Inzunza$^{a}$ and V\'i{}t Jakubsk\'y$^b$ 
}
\\[6mm]

\noindent ${}^a${\em Departamento de F\'isica\\ Universidad de Santiago de Chile,  Av. Victor Jara 3493, Santiago, Chile}\\[3mm]
\noindent ${}^b${\em
Nuclear Physics Institute\\ Czech Academy of Science, 250 68 \v{R}e\v{z}, Czech Republic}
\vspace{12mm}
\end{center}

\begin{abstract}
We develop a suitable technique to design zero-energy graphene models with radial electrostatic potentials capable of achieving electrostatic confinement. Using the Gaussian law for electrostatics, we derive the charge density associated with these potentials that correspond to concentric electrostatic rings. The technique is based on a modified supersymmetric transformation that allows to design time-reversal invariant interaction terms and to find the corresponding zero-energy bound states in analytical form. Consequently, solutions with the same probability density but different angular momentum are characterized by circular probability currents flowing in opposite directions.  The energies of the systems defined in two Dirac valleys (one-valley) have a fourfold (twofold) degeneracy. As an example of the technique, we construct a ring-decorated Coulomb potential that exhibits zero energy collapse and bound states together.

\end{abstract}

\section{Introduction}

Due to its remarkable physical properties, graphene is a promising material for the construction of new electronic devices. Graphene configurations that would have been considered unrealistic a decade ago are now possible due to the rapid development of experimental methods. Such systems can be $p-$ or $n-$doped using electronic and chemical methods \cite{Drag}, or mechanically deformed \cite{manes,Pablits},  allowing the charge density on its surface to be adjusted by applying external positive and negative voltages. These techniques have been successfully used to build structures such as graphene quantum dots \cite{JLee,LiLin,Yung}, circular $p-n$ junctions \cite{Yuhang,Fris}, and quantum rings \cite{Russo} in laboratory settings. In the current scenario, new theoretical models describing electronic properties of graphene are desirable.

An interesting example is the quantum ring (QR). It is a conducting ring-shaped region in nanoscale space with a local density of states of the same shape \cite{Fomin,Vief}. The first graphene-based QR was produced experimentally in order to investigate Aharonov-Bohm oscillations and the existence of persistent circular currents \cite{Russo,Neto,Wakker}. Nowadays,  systems such as ``whispering galleries" have been developed in the laboratory, being efficient traps for charge carriers within the material \cite{Zhao,Brun}. Analytically tractable models of electrostatic QR in graphene were studied theoretically, e.g., in \cite{Zarenia1,Zarenia2,Hewageegana,Downing,Ring1,BeSaGr}. It has been shown that these structures can provide confinement of quasi-particles of zero energy, overcoming the Klein tunneling that affects their localization \cite{Kat}.

So far, the models studied in the literature are relatively simple, but not very different among them, representing a single ring with either smooth or rectangular profile. To the best of our knowledge, theoretical models analyzing the physics of more complex structures such as concentric rings have not yet been discussed. The current article aims to address this gap by providing a robust framework for the theoretical design of a wide variety of radially symmetric electric potential enabling quantum rings, hosted by a single layer of graphene. Our approach is based on an analytical scheme known as supersymmetric transformations. 

The supersymmetric transformations plays an important role in the theory of integrable systems \cite{Matveev} and supersymmetric quantum mechanics \cite{Cooper}. The transformations (also called Darboux transformations in the literature) essentially map solutions of an initial differential equation into solutions of another equation of the same type, but with a modified potential term \cite{Darboux}. They found wide application in quantum mechanics when applied in the context of Schr\"odinger \cite{Darboux,Cooper,Junker,Matveev} or Dirac-type equations \cite{Samsonov,Samsonov2,Schulze1}, but also include optical and lattice systems, see e.g. \cite{re1,re2}. The transformation is represented by a differential operator uniquely defined by its kernel, which consists of a set of eigenstates of the initial Hamiltonian, usually called seed states. These operators connect the initial system with the new one in the so-called intertwining relation. In the same way, the seed states define the additional potential term of the new system. Due to the non-trivial nature of the operator kernel, the transformation map is not always one-to-one and, more generally, sometimes does not map states into physical ones. Therefore, the systems may have new spectral properties compared to the original. For example, the new model can have additional bound states, usually called \textit{``missing"} states in the literature.

The supersymmetric transformations were discussed both for stationary and non-stationary one-dimensional Dirac equations, see  \cite{Samsonov} and \cite{Samsonov2}, respectively. They have been used, for example, to reveal the connection between Klein tunneling in one-dimensional carbon structures and the free particle due to supersymmetry \cite{JaNiPl}, 
in the spectral design of twisted carbon nanotubes \cite{Jakubsky1}, in Dirac systems with transparent potentials \cite{Correa} and  systems described by the non-Hermitian Dirac equation \cite{Correa2}. They have also been generalized to higher spin operators in \cite{PozSchul}, representing an effective tool for studying a wide range of Dirac materials \cite{JakZel1,JakZel2}, see also \cite{Dmat1,Dmat2,Dmat3,BetKlein}. Although the results for higher spatial dimensions are limited \cite{Schulze2,asym1,Ioffe,asym2,CorInzJak}, and some of them rely on the adaptation of techniques originally developed for 1D or (1+1)D systems, they have been useful in revealing interesting physics of Dirac fermions in graphene. To give just a few examples, some of these approaches have been used to construct models that exhibit omnidirectional Klein tunnelling \cite{asym1} and, more recently, to explain the confinement in Lorentzian well potentials due to shape invariance \cite{CorInzJak}. Inspired by \cite{CorInzJak}, in this article we present a modified supersymmetric transformation based on the asymmetric intertwining relation. It allows to construct time-reversal invariant planar models of Dirac fermions in graphene with electrostatic potentials exhibiting rotational symmetry, while providing analytical expressions for their zero energy states. The generalization of the technique from one to two dimensions allows to study applied scenarios such as graphene. However, the main difference between the two cases is that the 2D case does not allow to map known states to new ones, mainly due to the fact that the intertwining relation is no longer symmetric. 

To demonstrate the applicability of the new technique, we modify the model of a Coulomb impurity in graphene. As a result, new systems with concentric rings surrounding the impurity are constructed. The new potentials do not prevent the fall to the center of the Dirac particles \cite{Colapse,Colapse2,QRig,colapselab}, but each of the rings is able to trap an electron-hole pair of zero energy with a determinate angular momentum quantum number that is intrinsically related to the geometry of the ring.

The article is organized as follows. In section \ref{secZeroE} we briefly review the physics of zero energy states for rotationally symmetric electrostatic potentials. The electrostatic confinement effect due to the presence of bound states is discussed, as well as a general perspective on atomic collapse phenomena for singular at the origin potentials. In section \ref{SecDefPot}, the modified supersymmetric transformation is presented, which enables construction of the new deformed systems with QR-type electric potentials. We show how the technique is based in a specific non-unitary transformation and asymmetric intertwining relations. In section \ref{SecRingDef} we use the technique to construct a Coulomb potential decorated by concentric rings.  We show that each of these rings supports a pair of zero energy bound states representing two confined carriers with opposite total angular momentum. It is also shown that the system exhibits the simultaneous existence of bound states and atomic collapse phenomena. In addition, physical aspects of the new potentials are discussed, in particular the effective charge density corresponding to the electric potential is calculated. Finally, we leave the section \ref{SecDis} for discussion and outlook. 

\section{Electrostatic confinement thought zero energy states}
\label{secZeroE}

 Electrostatic confinement in a graphene layer is of theoretical and experimental interest and is related to the possibility of creating bound states by the action of an electric field. Numerous studies have suggested that the Klein tunneling effect poses a challenge to achieving this phenomenon. Nevertheless, theoretical research has shown that certain electric field configurations, such as quantum dots\cite{Downing}, rings\cite{Ring1} and circular $p{-}n$ junctions\cite{pnjun}, can still achieve confinement for carriers with zero energy.  The experimental evidence also supports these findings, see for example \cite{JLee,LiLin,Yung}. 

This section analyzes some general aspects of the continuous model of graphene when an external radially symmetric potential is considered. We first examine the symmetries of the two-valley model and their corresponding reductions to the spin-$\frac{1}{2}$ case to emphasize the importance of Kramer's theorem, which explains the energy level degeneracy. Hereafter, we explore the electrostatic confinement effect more deeply by studying the zero energy equation and the conditions for creating zero energy bound states. Additionally, we investigate the possibility of having collapsing states instead of bound states when the potentials are singular at the origin of coordinates. Finally, we reconsider the renowned Coulomb problem from the perspective of zero energy states, demonstrating the impossibility of hosting bound states and the presence of the atomic collapse phenomenon.
 
\subsection{Symmetries of a graphene layer in presence of radial electric potentials}
Ideal graphene is described as a two-dimensional honeycomb carbon lattice \cite{Kat}.
The tight-binding approximation reveals that there are two inequivalent Dirac points (denoted $\mathbf{K}$ and $\mathbf{K'}\equiv-\mathbf{K}$) located in two, mutually opposite, corners of the first Brillouin zone. In the low energy regime, the charge carriers behave like Dirac fermions. They are described by the following equation,
\begin{equation}\label{DiracEqH}
\hbar v_F \mathbb{H}\Psi(\vr,t)=\Ti\hbar \partial_t\Psi(\vr,t)\,,\qquad \mathbb{H}=\begin{pmatrix}H_{\mathbf{K}}&0\\0&H_{\mathbf{K'}}\end{pmatrix},\qquad \Psi(\vr,t)=\begin{pmatrix}\psi_{\vK}(\vr,t)\\\psi_{\vK'}(\vr,t)\end{pmatrix}\,.
\end{equation}
Here, $H_\vK$ (or $H_{\vK'}$) is the Dirac Hamiltonian capturing dynamics in the vicinity of the Dirac point $\vK$ (or $\vK'$). Weyl spinor $\psi_\vK$ (and $\psi_{\vK'}$) represents Dirac fermion with its momentum being close to $\vK$ (or $\vK'$). For the external, radial symmetric potential interaction, the Hamiltonians read as
 \begin{equation}
H_\mathbf{K}=-\Ti\vsigma\cdot\nabla+V(r)\,, \qquad V(r)=\frac{e \phi(r)}{\hbar v_F}\,,
\qquad
 H_{\mathbf{K}'}=H_{{\mathbf{K}}}^T=\sigma_1 H_{{\mathbf{K}}}\sigma_1\,.
\end{equation}
Here,  $\vsigma=(\sigma_1,\sigma_2)$ with $\sigma_i$ denoting the Pauli matrices, $v_F$ is the Fermi velocity and $e$ is the electron charge. The effect of a scalar potential does not introduce inter-valley interaction but it can modify the energy spectrum of the system. This allows for the creation of bound states at zero energy for certain configurations, as we show below.

Equation (\ref{DiracEqH}) is symmetric with respect to the following set of transformations
\begin{align}
\label{T1}
&\Psi(\vr,t)\rightarrow \mathbb{T}_1\Psi(\vr,-t)& &\mathbb{T}_1= \sigma_2\otimes (\sigma_1                                                                                                            \mathcal{T})\\
\label{T2}
&\Psi(\vr,t)\rightarrow \mathbb{T}_2\Psi(\vr,-t)& &\mathbb{T}_2=\sigma_3\otimes \mathcal{T}\\
\label{R}
&\Psi(\vr,t)\rightarrow \mathbb{P}\Psi(\vr,t)&&\mathbb{P}=-\mathbb{T}_1\mathbb{T}_2=\sigma_1\otimes \sigma_1\\
\label{rot}
&\Psi (\vr,t)\rightarrow e^{\Ti \theta\mathbb{J}}\Psi(\vr,t) &&  \mathbb{J}=\begin{pmatrix}J&0\\0&\sigma_1J\sigma_1\end{pmatrix}
\end{align}
where $\mathcal{T}$ and $J$ are the spin-$\frac{1}{2}$ versions of the time reversal transformation and the generator of rotations \cite{Kat}. They are given by
\begin{equation}
\label{Spin1/2sym}
\mathcal{T}=T\sigma_2\,,\qquad T \Ti=-\Ti T\,,\qquad
J=-\Ti(x\partial_y-y\partial_x)+\frac{1}{2}\sigma_3\,.
\end{equation}
The transformation (\ref{T1}) represents the standard time reversal transformation for graphene, which swaps
the valley index but preserves the sublattice index of the Weyl spinors $\psi_\vK$ and $\psi_{\vK'}$ \cite{Kat}. 
 Similarly, we refer to (\ref{T2}) as the ``pseudo time reversal transformation," as it is antilinear but preserves the valley index.  Transformation (\ref{R}) swaps both the valley and pseudo-spin indices, yet, it is linear contrary to antilinear symmetries (\ref{T1}) and (\ref{T2}). Finally, (\ref{rot}) corresponds to spatial rotations by an angle $\theta$.
 It is worth noting that the set of discrete transformations
$\{\mathbb{I},\Ti\mathbb{T}_1,\Ti\mathbb{T}_2,\mathbb{P}\}$
forms a Klein-four group $\mathbb{Z}_2\cross \mathbb{Z}_2$ \cite{Klein4}. Additionally we have
\begin{equation}
\{\mathbb{J},\mathbb{T}_1\}=\{\mathbb{J},\mathbb{T}_2\}=[\mathbb{J},\mathbb{P}]=0\,,
\end{equation}
where $\{,\}$ denotes anti-commutator. These relations mean that $\mathbb{T}_1$ and $\mathbb{T}_2$ flip the sign of the total angular momentum of the states, while $\mathbb{P}$ do not affect this quantity. Additionally,
it is clear that for a given energy eigenvalue there exist four independent solutions
\begin{equation}
\Psi(\vr),\quad\mathbb{T}_1\Psi(\vr),\quad\mathbb{T}_2\Psi(\vr),\quad \mathbb{P}\Psi(\vr).
\end{equation}
In consequence, the system has a four-fold degeneracy.

As the operators $H_\vK$ and $H_{\vK'}$ are unitary equivalent, we can consider the dynamics in the vicinity of $\vK$ described by spin-$\frac{1}{2}$ Hamiltonian \footnote{We neglect the spin-degree of freedom of the Dirac fermions in graphene. We use ``spin" through out the article to refer to the pseudo-spin of the quasi-particles that distinguishes the two triangular sublattices of the hexagonal crystal.} $H_{\vK}$ without loss of generality,
 \begin{equation}\label{DiracEq}
\hbar v_F H\psi(\vr,t)=\Ti\hbar \partial_t\psi(\vr,t)\,,\qquad H\equiv H_\vK\,,\qquad \psi(\vr,t)\equiv \psi_\vK(\vr,t)\,.
\end{equation}
Within this scheme, the relevant symmetry transformations are just (\ref{Spin1/2sym}).
 
The Kramer's theorem implies that energy levels of spin-$\frac{1}{2}$ systems with time-reversal symmetry always exhibit even degeneracy \cite{KramersP,Landau}.  This conclusion can be made directly as for any solution of the stationary equation    
 \begin{equation}
\label{DiracEq}
\hbar v_F H \psi(\vr)=E \psi(\vr)\,,
\end{equation}
there is a second linearly independent solution $\mathcal{T}\psi(\vr)$. These two solutions are usually referred to as the Kramer pair in the literature. It plays an important role in the study of the spin Hall effect \cite{KramersP2}. The time-reversal symmetry changes sign of angular momentum and of the matrix vector $\vsigma$,
\begin{equation}
\label{tirev}
\mathcal{T} J\mathcal{T}^{\dagger}=-J\,, \qquad
\mathcal{T} \vsigma \mathcal{T}^{\dagger}=-\vsigma \, .
\end{equation}
These transformations mean that the spinors $\psi(\vr)$ and $\mathcal{T}\psi(\vr)$ transform under rotation with opposite orientations, and additionally the corresponding probability currents flow with opposite directions,      
\begin{equation}
\label{probcur}
\vj^+=\psi^\dagger \vsigma \psi\,,\qquad
\vj^-=(\mathcal{T}\psi)^\dagger \vsigma (\mathcal{T}\psi)=\psi^\dagger (\mathcal{T}^\dagger \vsigma \mathcal{T})\psi=-\vj^+\,.
\end{equation}
\subsection{The zero energy equation}
In the following, we address the electrostatic confinement by examining the equation (\ref{DiracEq}) for $E=0$ in polar coordinates,
 \begin{equation}
\label{initialPor}
H\psi=0\,,\qquad
H= \left(\begin{array}{cc}
V(r) & -\Ti e^{-\Ti \varphi}(\partial_{r}-\frac{\Ti}{r}\partial_\varphi)\\
 -\Ti e^{\Ti \varphi}(\partial_{r}+\frac{\Ti}{r}\partial_\varphi )& V(r)
 \end{array}\right)\,.
\end{equation}
To solve the equation, it is convenient to diagonalize $H$ and $J$ simultaneously, so that we consider states of the form  
\begin{equation}
\label{spinorm}
\psi_{m}=  e^{\Ti m \varphi } \left(
\begin{array}{c}
 \chi_m(r) \\
 \Ti e^{\Ti \varphi }\xi_m (r) \\
\end{array}
\right)\,,\qquad m=0,1,2,\ldots\,.
\end{equation}
It is sufficient to take only positive values for $m$, since due to (\ref{tirev})
we have
\begin{equation}
J\psi_{m}=\left(m+\frac{1}{2}\right)\psi_{m}\,,\qquad
J(\mathcal{T}\psi_{m})=-\left(m+\frac{1}{2}\right)(\mathcal{T}\psi_{m})\, ,
\end{equation}
i.e. the two states from the Kramer pair have opposite angular momentum. As should be expected, these states are orthogonal to each other, i.e., $\bra{\psi_m}\ket{\mathcal{T}\psi_m}=0$.  

Inserting the spinor ansatz (\ref{spinorm}) into the equation (\ref{initialPor}), two coupled equations of the radial functions  $ \chi_m(r) $ and $ \xi_m(r) $ are obtained,
\begin{equation}
\label{derivr}
 \frac{d \chi_m(r)}{dr}=\frac{m}{r} \chi_m(r) +V(r) \xi_m(r)\,,\qquad
\frac{d \xi_m(r)}{dr} =-\frac{m+1}{r}\xi_m(r)-V(r) \chi_m(r)\,.
\end{equation}
The equations in (\ref{derivr}) can also be decoupled as follows,
\begin{align}
\label{Radial}
&\left[\frac{d^2}{dr^2}+\left(\frac{1}{r}-\frac{1}{V(r)}\frac{d V(r)}{dr}\right)\frac{d}{dr} +V(r)^2-\frac{m^2}{r^2}+\frac{m }{r V(r)}\frac{d V(r)}{dr}\right]\chi_m(r)=0\,,
\\
\label{radxi}
&\xi_m(r)=\frac{1}{V(r)}\left( \frac{d}{dr}-\frac{m}{r}  \right)\chi_m(r)\,.
\end{align}

In order to make the discussion of the electric confinement more specific, it is necessary to provide more details on the potential and the boundary conditions. In this context, we assume that the graphene sheet is sufficiently large so that the electric field does not affect its boundaries as it has rather local effect. A good approximation for this situation is given by a system in the infinite plane with the potential with the following asymptotics,
\begin{align}
\label{Vaps}
r\rightarrow \infty\,,\qquad V(r)\sim -\frac{\beta}{r^n}\sim 0\,,\qquad n\geq 1\,,\qquad
[\beta]=[L]^{n-1}\,.
\end{align}
The bound states are required to be normalizable. Therefore, the  probability density
\begin{equation}
\rho_m(r)=\psi_m^\dagger \psi_m=(\mathcal{T}\psi_m)^\dagger(\mathcal{T}\psi_m)=|\chi_m(r)|^2+|\xi_m(r)|^2,
\end{equation}
should be integrable in the entire plane,  i.e.,
\begin{equation}
\label{Normalization}
\bra{\psi_m}\ket{\psi_m}=\bra{\mathcal{T}\psi_m}\ket{\mathcal{T}\psi_m}=2\pi \int_0^{\infty}r\rho(r)dr
<\infty\,.
\end{equation}
We prescribe the following boundary conditions in the origin,
\begin{equation}
\label{bc}
\chi_m(0)=\xi_m(0)=0\,,\qquad
\chi_m(r)_{r\rightarrow \infty}\sim r^{-a}\,,\qquad
\xi_m(r)_{r\rightarrow \infty}\sim r^{-b}\,,
\end{equation}
where we define $a,b>1$ to avoid logarithmic divergence of $\rho_m$.

On the same level of generality, we can compute the probability density currents (\ref{probcur}) for a state of the form (\ref{spinorm}) and its Kramer pair, yielding
\begin{equation}
\vj_m^\pm=\pm \psi_m^\dagger \vsigma \psi_m=N^2(|\chi_m|\xi_m+\chi_m|\xi_m|)\vvarphi, \qquad
\vvarphi=(-\sin\varphi,\cos\varphi)\,,
\end{equation}
where $N$ is the normalization constant.
The localized circular currents associated with the bound states $\psi_m$ or $\mathcal{T}\psi_m$ can create a magnetization in the material via spin-polarization mechanisms  \cite{Wakker,spintronic}.

Let us see whether it is possible to get solutions with the asymptotic behavior (\ref{bc}).
We take the long-range approximation of the equation (\ref{Radial}), having in mind the asymptotic form of the potential (\ref{Vaps}). We get
\begin{equation}
\label{asprad}
\left[\frac{d^2}{d r^2}+\frac{(n+1) }{r}\frac{d}{dr}+\frac{\beta^2}{r^{2n}}-\frac{m (m+n)}{r^2}\right]\chi_m (r) =0\,.
\end{equation}
By solving this equation we deduce the asymptotic behavior of the radial functions, which corresponds to
\begin{align}
\label{chixiasp1}
&\chi_m(r)_{r\rightarrow \infty}\sim \left\{ \begin{array}{lc}
c_1 r^{m} +\frac{c_2}{r^{m+n}} & n>1\\
c_1 r^{\gamma-1/2} +c_2 r^{-\gamma-1/2} & n=1
\end{array}
\right.
\,,\\
&\xi_m(r)_{r\rightarrow \infty}\sim
\left\{\begin{array}{lc}
 \frac{c_2}{r^{m+1}} & n> 1\\
 c_1\frac{(2m+1 -2\gamma)}{2\beta}r^{\gamma-1/2}+c_2\frac{(2m+1 +2\gamma)}{2\beta}r^{-\gamma-1/2}& n=1
\end{array}\right.\,, \label{chixiasp2}
\end{align}
Here, the dimensionless parameter $\gamma$ is defined as
\begin{equation}
\label{gamma}
\gamma=\gamma(m,\beta)=\frac{1}{2}\sqrt{(2m+1)^2-4\beta^2} \,.
\end{equation}
It can be identified with the celebrated \emph{``atomic collapse parameter"} and it will be discussed in more detail later in the text.
 
Revising the equation (\ref{chixiasp1}) and (\ref{chixiasp2}), it reveals that we can achieve the requested asymptotics of  bound states at infinity  (\ref{bc}) as long as  we fix $c_1=0$,  $\gamma>1/2$ , and $m>0$. When $\gamma$ is a pure imaginary, the functions $\xi_m$ and $\chi_m$ are oscillating and decay as $r^{-\frac{1}{2}}$. Therefore, they refrain from being square integrable and rather represent scattering states.

The boundary conditions at $r=0$ can be always achieved when the potential is regular at that point. In case of singular at the origin potentials, the well known atomic collapse phenomenon (also called falling to the center problem) can emerge, see e.g. \cite{Landau,singularP}. It occurs in electrostatic potentials that decay as $1/r^q$ with $q\geq 2$ in the non-relativistic case. It was discussed for Dirac particles in graphene in \cite{Kat,Colapse,Colapse2,QRig},  and addressed from the point of view of the renormalization group approach in \cite{GoPo}. Let us review the phenomenon here briefly, showing that relativistic particles  can also fall to the force center in the potential
\begin{equation}
V(r)\sim -\frac{\beta}{r}\, \quad \mbox{for} \quad  r\rightarrow 0.
\end{equation}
Indeed, the radial equation (\ref{Radial}) near $r=0$  and its solutions coincide with  (\ref{asprad}) and (\ref{chixiasp1}) for $n=1$.  
The parameter $\gamma$ separates the solutions into two different regimes that have qualitatively different physical implications. When $\gamma\in \R$, the centrifugal barrier in (\ref{asprad}) is dominant and the system is in the subcritical regime. In such a case, the boundary condition for bound states can be achieved by fixing $c_2=0$. The case of pure imaginary $\gamma$ corresponds to the critical regime, where the potential dominates over the centrifugal barrier and can attract the particle to the origin. In this case, the solutions will be singular in $r=0$, presenting pathological oscillations as $r\rightarrow 0$. Their explicit form is
 \begin{align}
 \label{apchia}
&\chi_{m}(r)|_{r\rightarrow 0}\sim c_1 r^{\Ti\omega-\frac{1}{2}}+c_2r^{-\Ti\omega-\frac{1}{2}}\,,\qquad
\omega= \frac{1}{2}\sqrt{4\beta^2-(2m+1)^2}
\,.
 \end{align}
 
Despite their bizarre nature, these solutions still carry crucial information about the model. For a deeper analysis, let us consider a regularized model by fixing the potential to be constant in the subregion $[0,r_0]$ with $r_0$  being near but different from zero. Subsequently, we study the limit $r_0\rightarrow 0$.

After matching the two parts of the solutions by using continuity conditions, we finally get
\begin{equation}
\label{finalAps}
 \chi_{m}\sim\left\{ \begin{array}{cc}
\frac{1}{\sqrt{r}}\cos(\omega\log(\frac{r}{r_0})+\theta) & r>r_0\\
C\frac{\sinh(kr)}{r} & 0<r<r_0
\end{array}
\right.\,,\qquad
k=\frac{1}{r_0}\sqrt{\omega^2+\frac{1}{4}}\,,
\end{equation}
where we have used the parameterization 
\begin{equation}
(c_1,c_2)=\frac{1}{2}(e^{-\Ti(\omega \ln(r_0)-\theta)},e^{\Ti(\omega \ln(r_0)-\theta)})\,,\qquad
\theta \in \R\,,
\end{equation}
and the constant $C$ is fixed by the requirement of continuity. When  $r_0\rightarrow0$, we can see how oscillations near the origin start to accumulate, increasing the number of zeros. This is the typical behavior of the solutions representing collapsing states in potentials that are too singular \cite{singularP,singularP2}. In one hand, the maximum of (\ref{finalAps}) is at $r_0$, near the force center, indicating the most probable position of the particle. On the other hand, semi-classical trajectories indicate a falling to the center of the particle in spiral paths \cite{QRig}. 

\subsection{The zero energy solutions in the Coulomb problem
\label{2.3}}
The Coulomb potential is the fundamental interaction that describes atoms in relativistic quantum mechanics, where $v_F\rightarrow c$. Unlike its non-relativistic counterpart, it allows for the possibility of atomic collapse \cite{Kat}. This effect can occur when $Z>\alpha^{-1}=137$, where $\alpha$ is the fine structure constant.  However, the prediction has not yet been confirmed due to the difficulty of producing heavy nuclei in the laboratory.  In graphene, the Coulomb interaction can be introduced as impurities in the honeycomb lattice \cite{Colapse, Colapse2}, and the threshold for observing atomic collapse is less difficult to reach. On the experimental side, there is evidence that STM microscopy can produce this effect in the laboratory \cite{colapselab}.

The Coulomb potential reads as
\begin{equation}
V(r)=-\frac{\beta}{r}\,,\qquad
\beta=\frac{1}{\hbar v_F }\frac{Ze^2}{\epsilon_{\text{ext}}} \,,
\end{equation}
where $\beta$ is the dimensionless potential strength, $\epsilon_{\text{ext}}$ is a dielectric constant related to an external substrate, and $Z$ is the atomic number. The corresponding zero energy solutions coincide with  (\ref{chixiasp1}) and (\ref{chixiasp2}) for $n=1$, but they are now valid for the entire space.

\begin{figure}[H]
\begin{center}
\includegraphics[scale=0.47]{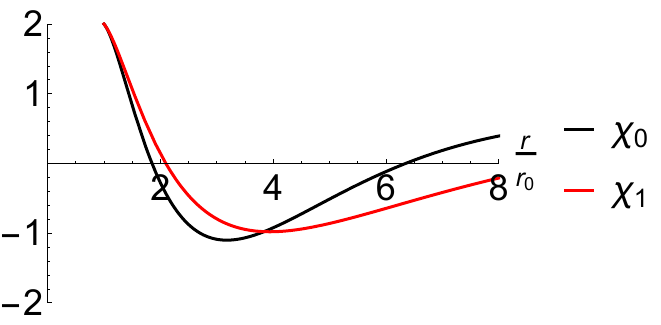}
\includegraphics[scale=0.47]{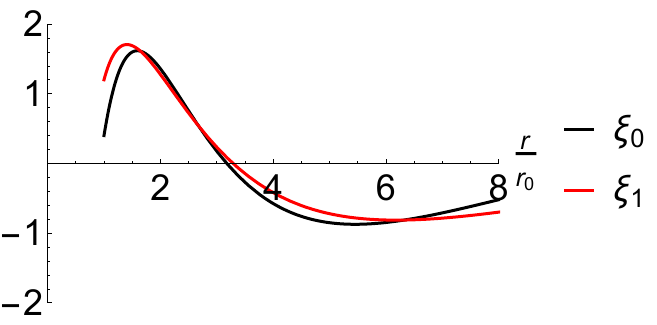}
\includegraphics[scale=0.47]{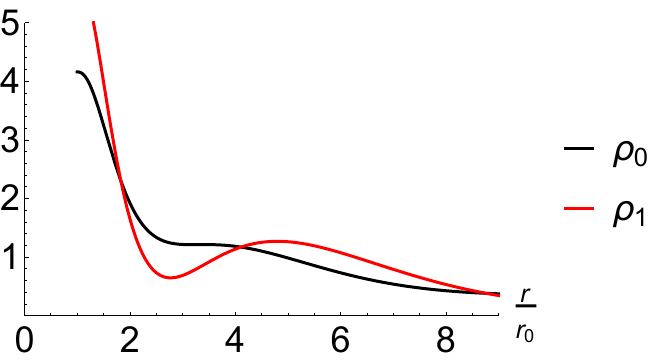}
\end{center}
\caption{\small{ The plots of the radial functions and the  (non-normalized) probability densities of the two collapse states when $\beta=\frac{5}{2}$ respect to $r/r_0$ are presented from left to write. The constant $r_0$ acts as a natural scale, and as we approach its value to zero, oscillations in the radial functions begin to accumulate near the force center. This behavior is typical of states in the critical regime in general. The probability density decreases slowly as the distance increases, but diverges like $1/r$ when approaching the origin.
When considering the regularized problem, one realizes that the particle is most probable to be found at $r=0$.}}
\label{colapse}
\end{figure}

For the solutions in the sub-critical regime, it is not possible to fix the integration constants such that the solution would fulfill the boundary conditions both at $r=0$ and $r\rightarrow+\infty$, see (\ref{bc}). Therefore, there are no bound states in the system. On the other hand, the solutions in the critical regime imply that the particle may fall to the center. In case of graphene, the collapsing states can occur whenever $\beta>\beta_c$, where the threshold value $\beta_c=\frac{1}{2}$ is defined as the potential strength for which $\gamma=0$ at $m=0$. The collapsing states are also characterized by having a relatively low angular momentum with $m$ in the interval $[0,\beta-\frac{1}{2})$. For example, fixing $\beta=5/2$, the collapsing states are those labeled by $m=0,1$. The corresponding radial functions and the probability density for such states are shown in the figure \ref{colapse}.

When considering energy levels other than zero, the solutions of the stationary equation should correspond to resonant quasi-bound states with complex energy \cite{QRig,GoPo}. These quasi-Rydberg states describe trapped electrons in a small region where the potential is regularized, acting as a barrier potential. The electrons escape from the barrier after a finite lifetime due to the Klein tunneling effect, providing partial confinement in the system. Nevertheless, absence of the localized bound states is disappointing since in the non-relativistic case, the system has discrete energies.

Concluding, Coulomb potential cannot host bound states. Nevertheless, it might not be excluded for the potentials that coincide with Coulomb at the origin and for large $r$, however, they differ in a rather localized region. In other words, there arises a natural question: can a local deformation of the Coulomb potential host the bound states?

In the following text, we provide an affirmative answer to the question. We develop an algorithm that allows for construction of the new potentials with exactly solvable zero energy solutions. These new potentials represent radially symmetric deformations of the known potentials that can host bound states. This formalism is applied to the Coulomb problem. The model is constructed where zero energy bound states and the collapsing states coexist together.

\section{Designing radial deformations of electrostatic potentials }
\label{SecDefPot}
In this section, the supersymmetric or Darboux transformation \cite{Cooper,Junker,Darboux,Matveev,Schulze1,Samsonov,Samsonov2} is adapted to create new radially deformed systems. In the usual setup, two differential operators, $h$ and $\tilde{h}$, are considered to be supersymmetric partners if there exists a differential mapping that transforms the eigenstates of one operator into the eigenstates of the other. This supersymmetric structure is encoded in the so-called intertwining relation
\begin{equation}
\label{Intertwining}
L h= \tilde{h}L\,,
\end{equation} 
 where $L$ is the intertwining operator. From this equation one automatically deduce the map 
 \begin{equation}
 h\psi=\lambda\psi\qquad\Rightarrow\qquad 
 \tilde{h}(L\psi)=\lambda (L\psi)\,.
 \end{equation}
This method has been widely used to construct and design one-dimensional solvable models describing non-relativistic and relativistic particles. 
It is assumed that all solutions of the operator $h$ are known. Some of them are used in explicit construction of $L$ and $\tilde{h}$.  In other words, starting from the initial system, we can create a new models using an algorithmic process. It is worth noticing that the Hamiltonians $h$ and $\tilde{h}$ do not need to be hermitian in general.

The standard supersymmetric transformation was proposed and widely used for one-dimensional systems. There are results on extension of the transformation to higher dimensional cases \cite{asym1,Schulze2}. Nevertheless, they do not provide such a generically applicable algorithm as in one dimension. 
Here we will employ a variant of the original method where the standard \emph{symmetric} intertwining relation (\ref{Intertwining}) will be replaced by an \emph{asymmetric} one $\tilde{L}h= \tilde{h}L$. This type of transformation has recently been explored in \cite{asym1,asym2,CorInzJak}. On one hand, the asymmetric intertwining relation is more limited as it only provides a mapping for a fixed energy level. On the other hand, it opens new possibilities, e.g. to analyze two-dimensional systems. 

Let us consider the operator (\ref{initialPor}). We multiply it by the factor $\Ti \sigma_2$ and consider such a Hamiltonian as the initial one,
\begin{equation}
\label{alternativePro}
h\psi=0\,,\qquad h=\Ti \sigma_2H=
\left(\begin{array}{cc}
e^{\Ti \varphi}( -\Ti \partial_r+\frac{1}{r}\partial_\varphi) & V(r)\\
-V(r)  &  e^{-\Ti \varphi}( \Ti \partial_r+\frac{1}{r}\partial_\varphi)
\end{array}\right)\,.
\end{equation}

The operator $h$ in (\ref{alternativePro}) is not hermitian. However, any operator such as this can be transformed into a Dirac Hamiltonian with an external electric field by multiplying it from the left by $-\Ti\sigma_2$,  as long as $V(r)$ is real. Although $H$ and $h$ do not commute, they share zero energy solutions and are time reversal invariant. The advantage of the working with $h$ instead of $H$ lies in the fact that it is covariant under the following supersymmetric  transformation $L$, see \cite{Matveev,Samsonov,Samsonov2},
\begin{align}
\label{DarLhS}
L=\cos\varphi \, \partial_r-\frac{1}{r}\sin\varphi \, \partial_\varphi-\Sigma\,,\qquad
\tilde{h}=h-\Ti[\sigma_3,\Sigma]\,,\qquad
\Sigma= \left(\cos\varphi \, \partial_r U -\frac{1}{r}\sin\varphi \, \partial_\varphi U\right)U^{-1}\,.
\end{align}
Here, the matrix $U$ is composed from zero modes $\psi_1$ and $\psi_2$ of $h$, $U=(\psi_1,\psi_2)$, $hU=0$. It is frequently called ``seed matrix" in the literature and satisfies $LU=0$ by construction. Different choices of $\psi_1$ and $\psi_2$ give rise to different $L$ as well as $\tilde{h}$. 

This freedom in choice of the seed matrix can be utilized to acquire  Dirac Hamiltonian with new electric fields whose zero energy solutions are known. 

To construct the desired matrix $U$, we consider an arbitrary Kramer pair for the seed solutions in the following way,
\begin{equation}
U_m=(\psi_m, \mathcal{T}\psi_m)=  \left(
\begin{array}{cc}
 e^{\Ti m \varphi }\chi_m & -e^{-\Ti (m+ 1)\varphi }\xi_m  \\
 \Ti e^{\Ti (m+ 1)\varphi }\xi_m  &  \Ti e^{-\Ti m \varphi }\chi_m\\
\end{array}
\right)\,.
\end{equation}
The radial functions $\chi_m=\chi_m(r)$ and $\xi_m=\xi_m(r)$ are fixed as real linear independent solutions of the equations (\ref{derivr}). A remarkable property of $U_m$ is that it satisfies the relation $\mathcal{T} U_m \mathcal{T}^\dagger=\Ti U_m$, which means that the matrix $\Sigma$ and the operator $\tilde{h}_m$ are time reversal invariant by construction. Explicitly, the newly constructed operator $\tilde{h}$ is given by  
\begin{equation}
\label{htilde}
\tilde{h}_m=
\left(\begin{array}{cc}
e^{\Ti \varphi}( -\Ti \partial_r+\frac{1}{r}\partial_\varphi) &  e^{-2 \Ti \varphi }V_m(r)\\
- e^{2 \Ti \varphi }V_m(r)  &  e^{-\Ti \varphi}( \Ti \partial_r+\frac{1}{r}\partial_\varphi)
\end{array}\right)\,,
\end{equation}
where $V_m(r)$ is obtained via (\ref{DarLhS}). Its final form is 
\begin{align}
\label{v(rr)}
&V_m(r)=V(r)+V_m^{\text{def}}(r)\,,\\
&V_m^{\text{def}}(r)=\frac{ \chi_m(r)\partial_r\xi_m(r)- \xi_m(r) \partial_r\chi_m(r)}{\chi_m^2+\xi_m^2}=-2\left(\frac{ (2 m+1)\chi_m\xi _m}{r(\chi_m^2+\xi_m^2)}+V(r)\right)\,,
\end{align}
Obviously, $V_m(r)$ represents a radially deformed $V(r)$.  When compared (\ref{alternativePro}), the potential term in (\ref{htilde}) is multiplied by phase factors that would prevent to make the inverse mapping of $\tilde{h}_m$ into the requested form of Dirac operator with electric field. Fortunately,  the complex phases can be eliminated by a non-unitary transformation generated by the following matrix
\begin{equation} 
S=\left(\begin{array}{cc} re^{\Ti \varphi} & 0\\ 0 & re^{-\Ti\varphi} \end{array}\right)\,
\end{equation} 
and its inverse. Thus, the transformed operator
\begin{equation}
h_m=S\tilde{h}_mS^{-1}= 
\left(\begin{array}{cc}
e^{\Ti \varphi}( -\Ti \partial_r+\frac{1}{r}\partial_\varphi) & V_m(r)\\
-V_m(r)  &  e^{-\Ti \varphi}( \Ti \partial_r+\frac{1}{r}\partial_\varphi)
\end{array}\right)
\end{equation}
is of the form (\ref{alternativePro}) with the change $V(r)\rightarrow V_m(r)$. Finally, we can turn the $h_m$ into the physical Hamiltonian by multiplying it by $-\Ti \sigma_2$ from the left,  
\begin{equation}
\label{Hm}
H_m=-\Ti\sigma_2h_m=\left(\begin{array}{cc}
V_m(r) & -\Ti e^{-\Ti \varphi}(\partial_r-\frac{\Ti}{r}\partial_\varphi)\\
 -\Ti e^{\Ti \varphi}(\partial_r+\frac{\Ti}{r}\partial_\varphi )& V_m(r)
 \end{array}\right)=-\Ti \vsigma \cdot \nabla+V_m(r)\,.
\end{equation}
In this way we have successfully created a new pseudo-spin $\frac{1}{2}$ system with a radially deformed scalar potential. The specific features of the deformations are determined by the seed functions $\chi_m$ and $\xi_m$ and can vary depending on the original potential. Since $\chi_m$ and $\xi_m$ were chosen as linearly independent functions, the term $\chi_m^2+\xi_m^2$ never vanishes in the range $(0,\infty)$ and the deformation is free of singularities. 

Once we have constructed the Hamiltonian (\ref{Hm}), we move on to the analysis of the zero energy solutions. To do this, we include $S$ in the intertwining relation (\ref{Intertwining}), which results in
\begin{equation}
\label{interfi}
\mathcal{L}_m h=h_m\mathcal{L}_m\,,\qquad
\mathcal{L}_m=SL_m\,,
\end{equation}
where the explicit form of the modified intertwining operator $\mathcal{L}_m$ is
\begin{align}
\label{InterLL}
\mathcal{L}_m=\left(
\begin{array}{cc}
e^{\Ti \varphi}(r\cos\varphi\,\partial_r-\sin\varphi\,\partial_\varphi)+
 \frac{(m+1) \xi_m ^2-m \chi_m ^2}{\xi_m^2+\chi_m^2} & \Ti (r \cos (\varphi ) V(r)+\frac{e^{-\Ti \varphi } (2 m+1) \xi_m \chi_m}{\xi_m^2+\chi_m^2}) \\
 \Ti(r \cos (\varphi ) V(r)+\frac{e^{\Ti \varphi } (2 m+1) \xi_m \chi_m}{\xi_m^2+\chi_m^2}) &e^{-\Ti \varphi}(r\cos\varphi\, \partial_r-\sin\varphi\, \partial_\varphi)+ \frac{(m+1) \xi_m^2-m \chi_m^2}{\xi_m^2+\chi_m^2} \\
\end{array}
\right)\,.&
\end{align}
The new transformation (\ref{InterLL}) is inspired in the recently found transformations \cite{CorInzJak} that allowed to connect the Lorentzian well with the free particle. If we now multiply (\ref{interfi}) by $\sigma_2$, we finally get to the \emph{asymmetric} intertwining relation mentioned above,
\begin{equation}
\label{ASyminter}
\sigma_2\mathcal{L}_m\sigma_2 H=H_m\mathcal{L}_m\,.
\end{equation}
This kind of intertwining relations allows only a mapping between the hermitian Hamiltonians for a fixed energy level, which in this case is at $E=0$,  
\begin{equation}
H\psi=0\qquad\Rightarrow \qquad
H_m(\mathcal{L}_m\psi)=0\,.
\end{equation}
The form of the zero energy states for the model (\ref{Hm}) is obtained explicitly in terms of the radial seed solutions by directly applying $\mathcal{L}_m$ to the states (\ref{spinorm}) and using equations (\ref{derivr}),
\begin{equation}
\label{transSol}
\psi_{m,l}=\mathcal{L}_{m}\psi_{l}=e^{\Ti \varphi l}\left(
\begin{array}{c}
\chi_{m,l}\\
 \Ti e^{\Ti \varphi}\xi_{m,l}
\end{array}
\right)\,,  \quad l\neq m\, ,
\end{equation}
where the new functions $\chi_{m,l}$ and $\xi_{m,l}$ are given by,
 \begin{align}
 \label{chimm}
\chi_{m,l}= 
\frac{(l+m+1) \xi _m^2+(l-m) \chi _m^2}{\xi _m(r){}^2+\chi _m(r){}^2}\chi _l(r) 
-\frac{(2 m+1)  \xi _m \chi _m}{\xi _m^2+\chi _m^2}\xi _l \, ,
\\
 \label{ximm}
\xi_{m,l}=\frac{(2 m+1)  \xi _m \chi _m}{\xi _m^2+\chi _m^2} \chi _l
-
\frac{ (l-m) \xi _m^2+(l+m+1) \chi_m^2}{\xi _m^2+\chi _m^2}\xi _l
\,.
 \end{align}
Remarkably, the transformation preserves the total angular momentum eigenvalue of the states. It can be understood from the commutator between $\mathcal{L}_m$ and $J$, 
\begin{equation}
\label{comLAJ}
[\mathcal{L}_m,J]=S\sigma_2H\,,\qquad
\end{equation}
which vanishes in the subspace of zero energy states. Consequently, the modified transformation allows us to obtain all eigenstates labelled $l\not=m$. The special case for $l=m$, which cannot be obtained by the described procedure, can be constructed from the hermitian conjugation of the asymmetric intertwining relation (\ref{ASyminter}), which gives  
\begin{equation}
\label{ASyminter2}
\sigma_2\mathcal{A}_m\sigma_2 H_m=H\mathcal{A}_m\,,\qquad
\mathcal{A}_m=-\sigma_2\mathcal{L}_m^\dagger\sigma_2\,.
\end{equation}
By construction, the above asymmetric relation represents the transformation that maps the states of 
$H_m$ into the states of $H$ using
 $\mathcal{A}_m$,
\begin{equation}
 H_m\psi_{m,l}=0\qquad\Rightarrow \qquad
H(\mathcal{A}_m\psi_{m,l})=0\,.
\end{equation}
It represents the inverse action when compared to $\mathcal{L}_m$ and its intertwining relation (\ref{ASyminter}). The operator (\ref{ASyminter2}) also admits the representation $\mathcal{A}_m=SA_m$, where $A_m$ is an operator 
of similar form to $L$ in (\ref{DarLhS}), but with a different matrix $U\rightarrow \tilde{U}_m $, given by
\begin{equation}
\label{missingstates}
\tilde{U}_m = -\sigma_2 (S^\dagger)^{-1}(U_m^{-1})^\dagger = (\psi_{m,m},T\psi_{m,m})\,,\qquad
\psi_{m,m} =\frac{e^{\Ti m \varphi } }{r(\chi_m^2+\xi_m^2)}\left(
\begin{array}{c}\xi_m (r) \\
 \Ti e^{\Ti \varphi } \chi_m (r) \\
\end{array}
\right)\,.
\end{equation}
The Kramer pair of states $\psi_{m,m}$ and $T\psi_{m,m}$ are obviously zero energy solutions $H_m\psi_{m,m}=0$. They are the missing states that complement the set of solutions (\ref{transSol}). This way, the task of finding the full set of zero energy eigenfunctions of the new model $H_m$ is completed. Interestingly, the missing states hide relevant physics depending on the choice of $\chi_m(r)$ and $\xi_m(r)$. To explore this, we compute their probability density and the current density, 
\begin{equation}
\rho_{m,m}=\psi_{m,m}^\dagger\psi_{m,m}=\frac{1}{r^2(\chi_m^2+\xi_m^2)}\,,\qquad
j_{m,m}=\psi_{m,m}^\dagger\vsigma \psi_{m,m}=\frac{\chi_m\xi_m}{r^2(\chi_m^2+\xi_m^2)^2}\vvarphi\,.
\end{equation}
These relations suggest that the missing states $\psi_{m,m}$ and $T\psi_{m,m}$ can be normalizable and, therefore, represent the bound states, provided that the radial seed functions $\chi_m$ and $\xi_m$ diverge for large $r$. This can be achieved by fixing the later functions as the diverging non-physical eigenstates of $H$.

The whole algorithm can be repeated, fixing $H_m$ as the initial Hamiltonian this time. Fixing a Kramer pair of zero energy eigenstates of $H_m$, $\psi_{m, l}$ and 
$\mathcal{T}\psi_{m, l}$, as the new seed solutions, the new system can be generated with the following deformed potential,
\begin{align}
\label{step2}
V_{m,l}(r)=V(r)+\frac{2(2m+1)\chi_m\xi_m}{r(\chi_m^2+\xi_m^2)}-\frac{2(2l+1)\chi_{m,l}\xi_{m,l}}{r(\chi_{m,l}^2+\xi_{m,l}^2)}\,.
\end{align}
Repeated application of the procedure can result in a wide variety of systems with deformed potentials and multiple zero energy bound states. In the next section, we will illustrate the framework on the construction of radial deformations of the Coulomb potential.

\section{Ring deformations of the Coulomb problem}
\label{SecRingDef}

In this section, we employ the formalism developed in section  \ref{SecDefPot} to generate the new model where both bound states and collapsing states coexist. It corresponds to local ring-type deformations of the Coulomb potential. 

\subsection{Single ring}
The first step is to fix the Coulomb Dirac Hamiltonian as our initial system. Next, we choose linearly independent seed radial functions by the appropriate selection of the integration constants in (\ref{chixiasp1}) and (\ref{chixiasp2}) for $n=1$. In particular, we fix
\begin{align}
\label{chiCou}
&\chi_{m}(r)=r^{\gamma -\frac{1}{2}}-\frac{(2 m+1-2 \gamma)  r_c^{2 \gamma }}{2 \beta } r^{-\gamma -\frac{1}{2}}\,,
\\
\label{xiCou}
&\xi_{m}(r)=\frac{2 m+1-2 \gamma}{2 \beta }\left[r^{\gamma -\frac{1}{2}}-\frac{(2 m+1+2 \gamma ) r_c^{2 \gamma }}{2 \beta }r^{-\gamma -\frac{1}{2}}\right]\,,
\end{align}
where $r_c$ is a constant of length dimension whose physical meaning will be discussed below. To avoid problems with self-adjoint extensions, we define  $\gamma(m, \beta)$ in (\ref{gamma}) to be 
\begin{equation}
\label{paramcondi}
\gamma>1\quad \Longrightarrow \quad
m(m+1)-\frac{3}{4}>\beta^2\,.
\end{equation}
In this way, inserting the functions (\ref{chiCou}) and (\ref{xiCou}) into the equation (\ref{v(rr)}), we obtain the deformed potential 
\begin{equation}
\label{paramdef}
V_{m}(r)=
-\frac{1} { r }\left[\beta  -
\frac{8 \gamma ^2 \left(r/r_c\right)^{2 \gamma }}{(2 m+1) \left(r/r_c\right)^{4 \gamma }-4 \beta  \left(r/r_c\right)^{2 \gamma }+2 m+1}
\right]\,.
\end{equation}
A simple inspection shows that the rational deformation term is just a local perturbation so that the potential converges to that of Coulomb at origin and at infinity. Furthermore, it can be shown that the denominator becomes zero only for the complex values  $r_\pm^{2\gamma}=r_c^{2\gamma}(2\beta\pm \Ti \gamma)(1+2m)^{-1}$ provided (\ref{paramcondi}). Due to the requirement of linear independence of the seed solutions, the potential does not introduce any new singularities on the positive real line. 

\begin{figure}[H]
\begin{center}
\includegraphics[scale=0.7]{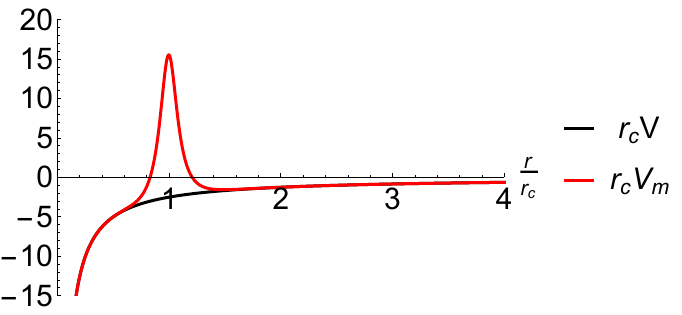}
\end{center}
\caption{\small{
The comparison between the original and deformed potentials as a function of $r/r_c$.
}}
\label{fig2}
\end{figure}

We can also identify some additional features that determine the nature of the deformation. For example, the potential (\ref{paramdef}) reaches its maximum value at $r=r_c$,
\begin{equation}
\label{vmmax}
V_{m}(r_c)=\frac{\beta+1+2m}{r_c}\,.
\end{equation}
Additionally, we can find the values of $r$ where the deformation and the Coulomb term cancel each other and the potential vanishes. This occurs at the following points,
\begin{equation}
\label{rinext}
r_1=  \left[\frac{2(\beta ^2+2 \gamma ^2 -\gamma  \sqrt{3 \beta ^2+4 \gamma ^2})}{(1+2m)\beta}
\right]^{\frac{1}{2\gamma}}
r_c\,,\qquad
r_2=\left[\frac{2(\beta ^2+2 \gamma ^2 +\gamma  \sqrt{3 \beta ^2+4 \gamma ^2})}{(1+2m)\beta}
\right]^{\frac{1}{2\gamma}}
r_c\,.
\end{equation}
Based on this analysis, we can conclude that the deformation forms a positive and smooth barrier in the general case. In Fig. \ref{fig2} we compare the original and the deformed potentials for the cases $\beta=\frac{5}{2}$ and $m=6$, where we can visualize the emergence of the barrier with maximum $r_c$. 

\vskip-0.25cm
The bound states of the deformed systems correspond to the Kramer pair $\psi_{m,m}$ and $\mathcal{T}\psi_{m,m}$, see  (\ref{missingstates}). Let us compute the corresponding the probability density and the current density, 
\begin{equation}
\label{rrhoCou}
\rho_{m,m}=\frac{N^2}{r_c^2} \frac{\eta ^{2 \gamma -1}}
{(2 m+1) \eta ^{4 \gamma }-4 \beta  \eta ^{2 \gamma }+2 m+1}\,,\quad
\vj_{m,m}= \frac{N^2}{r_c^2}\frac{\beta  \eta ^{4 \gamma }+\beta -(2 m+1) \eta ^{2 \gamma } }
{\left((2 m+1) \eta ^{4 \gamma }-4 \beta  \eta ^{2 \gamma }+2 m+1\right)^2}\eta ^{2 \gamma -1}\vvarphi\,. 
\end{equation}
Here, we introduced a dimensionless parameter $\eta=r/r_c $. 
The function $\rho_{m,m}(r)$ is regular in the entire plane as long as $\gamma>1$, vanishes at the origin and behaves at long distances as $\sim \eta^{-3\gamma}$. Since there are no singularities, the states composing the probability density are normalizable bound states. Here $N$ is a normalization constant. The density current has similar asymptotic properties. Additionally, it reflects circular motion of the particle. At $r_c$ it takes the following value
\begin{eqnarray}
\label{vjrc}
\vj_{m,m}(r_c)=-\frac{N}{r_c^2(4 +8 m-8 \beta)}\vvarphi\,.
\end{eqnarray}
Since (\ref{vjrc}) is negative, see (\ref{paramcondi}), the particle flow is clockwise. The functions (\ref{rrhoCou}) are plotted in figure \ref{fig3} for the case $\beta=5/2$ and $m=6$. 
\begin{figure}[H]
\begin{center}
\includegraphics[scale=0.8]{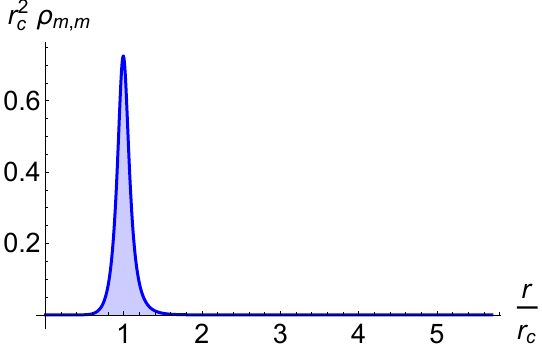}\,
\includegraphics[scale=0.4]{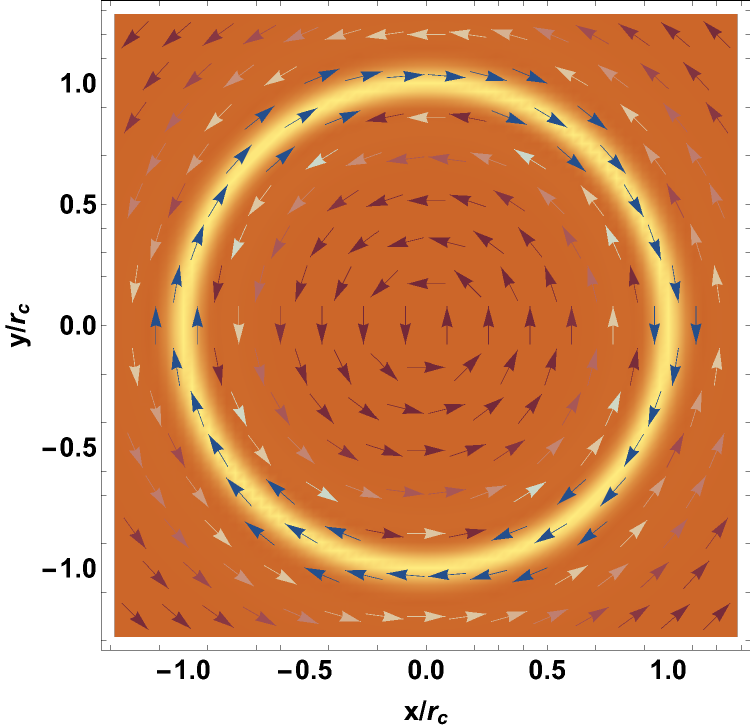}
\end{center}
\caption{\small{Plots of the normalized probability density (left) and the current vector density times $r_c^2$ (right).  The parameter values are $\beta=5/2$ and $m=6$. In this case, the numerical integration determines the normalization constant vale to be $N^2\approx 0.09$.}}
\label{fig3}
\end{figure}
\vskip-.25cm
To compute and analyse the remaining zero energy solutions, it is useful to rewrite the zero energy solutions of the Coulomb problem as follows 
\begin{equation}
\label{couldiv}
\chi_{\beta,l}(r;\pm \gamma')=r^{\pm \gamma'-\frac{1}{2}}\,,\qquad
\xi_{\beta,l}(r;\pm \gamma')=\frac{(2l+1 \mp 2\gamma')}{2\beta}r^{\pm\gamma-\frac{1}{2}}\,,
\end{equation}
where $\gamma'=\gamma(l)$. Then the formulas (\ref{chimm}) and (\ref{ximm}) can be employed, providing us with the following result  
\begin{align}
\label{chimmb}
\chi_{\beta,m,l}(\gamma')=f_-(\eta) \chi_{\beta,l}(\gamma')-f_0(\eta)\xi_{\beta,l}(\gamma')\,,\qquad
\xi_{\beta,m,l}(\gamma')=f_0(\eta) \xi_{\beta,l}(\gamma')-f_+(\eta)\chi_{\beta,l}(\gamma')\,,
\end{align}
where the functions $f_{i}(\eta)$ are given by
\begin{align}
f_\pm(\eta)=l +\frac{1}{2}\pm \frac{\gamma  (2 m+1) (\eta ^{4 \gamma }-1)}
{(2m+1)\eta ^{4 \gamma }-4 \beta  \eta ^{2 \gamma }+2 m +1}\,, \quad 
f_0(\eta)= \beta -\frac{4 \gamma ^2 \eta ^{2 \gamma }}
{(2 m+1) \eta ^{4 \gamma }-4 \beta  \eta ^{2 \gamma }+2 m+1}\,. 
\end{align}
The asymptotic form of the functions (\ref{chimmb}) near zero and at infinity correspond to 
\begin{align}
&\chi_{\beta,m,l}|_{\eta\rightarrow 0}=(\gamma+\gamma')r^{\frac{1}{2}+\gamma'}\,,\qquad
\xi_{\beta,m,l}|_{\eta\rightarrow 0}=(\gamma-\gamma')r^{\frac{1}{2}+\gamma'}\,,
\\
&
\chi_{\beta,m,l}|_{\eta\rightarrow \infty}=(\gamma-\gamma')r^{\frac{1}{2}-\gamma'}\,,\qquad
\xi_{\beta,m,l}|_{\eta\rightarrow \infty}=(\gamma+\gamma')r^{\frac{1}{2}-\gamma'}\,.
\end{align}
It means that the divergent states remain divergent after the transformation, while the collapsing states are transformed into collapsing states. This result is easy to understand due to the local nature of the deformation; the form of the interaction near zero and at infinity is just the Coulomb potential and, therefore, the analysis of subsection \ref{2.3} is still applicable. 
We show the density plots of the two collapsing states existing for the case $\beta=5/2$ and $m=6$ in figure \ref{colapse2}.
\begin{figure}[H]
\begin{center}
\includegraphics[scale=0.6]{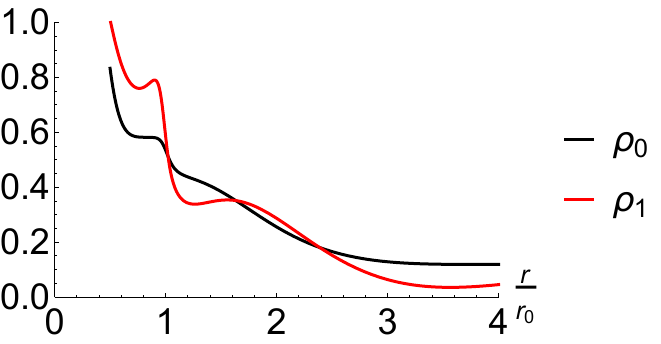}
\end{center}
\caption{\small{ Plots for the (not normalized) probability densities of the collapsing states in the case $\beta=5/2$ and $m=6$. For simplicity we take $r_c=1$, while $r_0$ has the same role as in figure \ref{colapse}.  They corresponds to real combination of the mapping of the collapsing states for the original model.
}}
\label{colapse2}
\end{figure}  

To conclude this paragraph, it is essential to note that if we replace the seed functions (\ref{chiCou})-(\ref{xiCou}) by the same combination but with the change $r^{2\gamma}\rightarrow -r^{2\gamma}$, the resulting potential deformation from the transformation will be a well instead of a barrier. Despite this difference, the analyses conducted for the barrier deformation can be applied, obtaining the analogous general results.

\subsection{Repeated application of the algorithm}

In this section, we shall illustrate how the algorithm for construction of the radial deformations can be applied repeatedly. We fix the system described by (\ref{paramdef}) as the initial one.  We take as the new seed radial functions the following combinations of (\ref{chimmb}),
\begin{align}
\chi_{m,l}= \chi_{\beta,m,l}(\gamma')+B r_c^{2\gamma'}\chi_{\beta,m,l}(-\gamma')\,,\qquad
\xi_{m,l}= \xi_{\beta,m,l}(\gamma')+B r_c^{2\gamma'}\xi_{\beta,m,l}(-\gamma')\,,
\end{align}
where $B$ is some real constant. The transformation then gives the potential (\ref{step2}). Instead of showing its rather complicated analytic form, we can compare its plots in figure \ref{fig3} with the Coulomb potential and with our initial barrier.
\begin{figure}[h!]
\begin{center}
\includegraphics[scale=0.55]{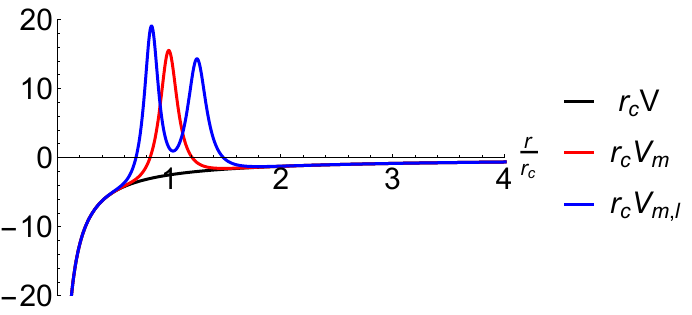}
\includegraphics[scale=0.55]{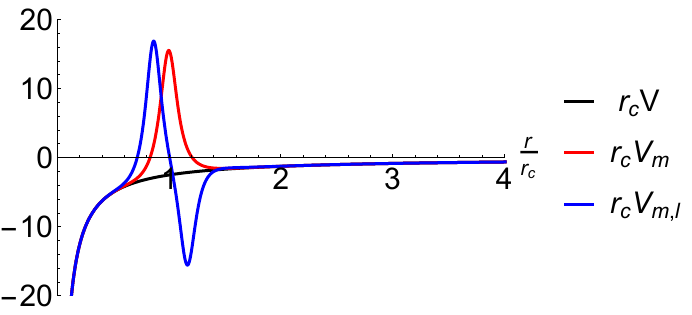}
\includegraphics[scale=0.55]{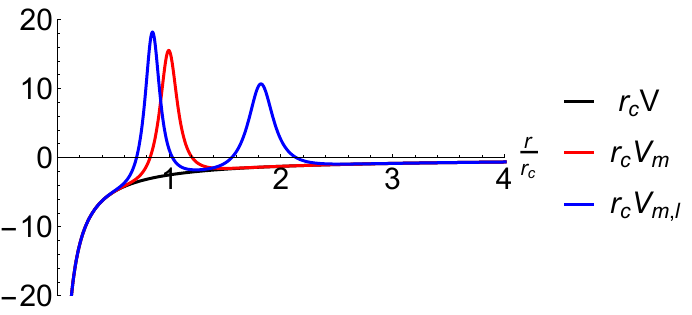}
\includegraphics[scale=0.55]{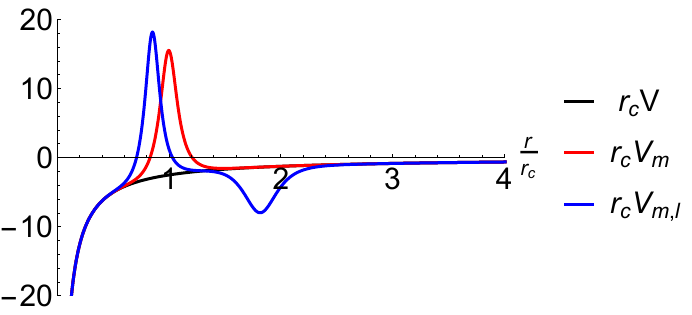}
\end{center}
\caption{\small{Different shapes of the potential $V_{l,m}$ depending on the values of $B$ for $\beta=5/2$, $m=6$ and $l=8$.
Left panel: $B=1/2$ (top) and $B=400$ (bottom). Right panel: $B=-1/2$ (top) and $B=-400$ (bottom).}}
\label{fig3}
\end{figure}
Obviously, the sign of $B$ determines whether the transformation introduces a barrier or a well potential. On the other hand, the magnitude of this parameter controls the distance where the new deformation is located.

\begin{figure}[h!]
\begin{center}
\includegraphics[scale=0.55]{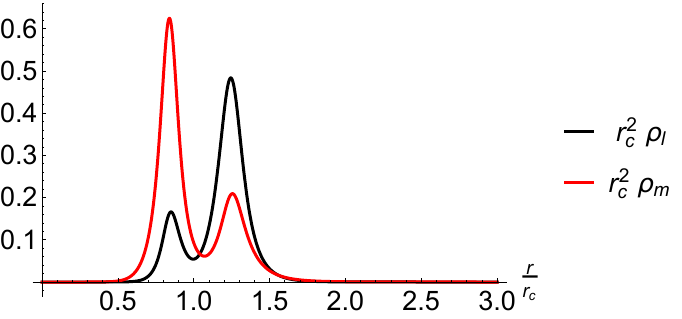}\qquad
\includegraphics[scale=0.55]{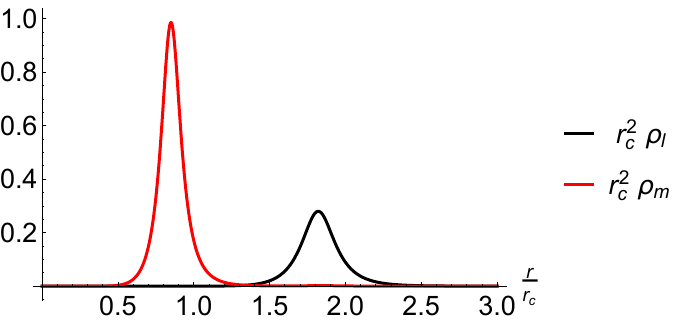}
\end{center}
\caption{\small{The figures correspond to the density plots for the cases $B=1/2$ (left) and $B=400$ (right), respectively. For small $B$, the potentials are close to each other and the particle can be found in both deformations. For large $B$, the distance between the potentials is larger and the particles are localized in their respective traps.}}
\label{fig6}
\end{figure}

The bound states forming two Kramer pair, corresponding to the angular momentum $m$ and $l$, can be found. The first pair comes from the direct mapping of the bound states of the initial systems. The second one can be found via the formula (\ref{missingstates}) for the missing states. The figure \ref{fig6} shows the position where the particles are most likely to be found, which somehow coincides with the positions of the maximal or minimal values of the potential.

The full scheme of these results can be generalized as follows: Starting from the Coulomb potential, we choose $N$ seed radial functions of the form 
\begin{align}
&\chi_{m}= \chi_{\beta,m,l}(\gamma')+B_m r_c^{2\gamma'}\chi_{\beta,m}(-\gamma')\,,\quad
\xi_{m}= \xi_{\beta,m}(\gamma')+B_m r_c^{2\gamma'}\xi_{\beta,m}(-\gamma')\,,\\
& m\in\mathfrak{m}=\{m_1,\ldots, m_N\}\,.
\end{align}
In each of them, we can fix the multiplication constants $B_m$ differently in value and sign. 
 
\begin{figure}[H]
\begin{center}
\includegraphics[scale=0.75]{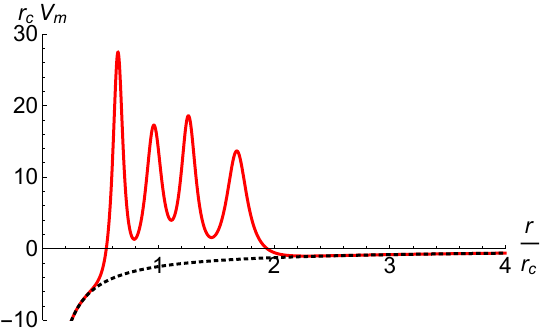}
\end{center}
\caption{\small{Deformation with four barrier potentials. Here we have used $\mathfrak{m}=\{6,8,10,12\}$ with $\beta=5/2$.}}
\label{fig1000}
\end{figure}

We start the process by using those with $m_1$ to produce the first transformation, resulting in the single barrier studied in this section. The next step is to map the second seed radial functions and use them to perform the second transformation. We repeat this procedure iteratively with all the seed functions in our set. This will result in a deformed potential $V_{\mathfrak{m}}(r)$ that has $N$ local deformations with a barrier or well shape, which forms and positions will be determinate bay de values and signs of constants $B_m$, see for instance figure \ref{fig1000}. Each deformation has a pair of bound states with opposite total angular momentum trapped on it, denoted by $m\in \mathfrak{m}$. 

\newpage

\subsection{Physical interpretations of the deformed potentials}
Effective electrostatic potential can be created in graphene in various ways. Just to mention a few, it can be produced by local deformations \cite{manes,Pablits}, 
electrical and chemical doping \cite{Drag}, by applying electrodes or just by putting the sample near a STM \cite{
JLee,LiLin}.  In each of this scenarios, the electric field can be associated with an effective charge density in the layer. Accordingly, the potentials constructed with the formalism explained in section \ref{SecDefPot}, which have the form 
$
\mathcal{V}(r)=V(r)+V^{\text{def}}(r)
$,
must satisfy the Poisson equation for electrostatics
\begin{equation}
\nabla^2\left(\frac{\hbar v_F}{e} \mathcal{V}\right)=\frac{\hbar v_F}{e} \left(\nabla^2V+\nabla^2V^{\text{def}}\right)=-\frac{1}{\epsilon}(n(\vr)+n^{\text{def}}(\vr))
\end{equation}
where $n(\vr)$ and $n^{\text{dif}}(\vr)$ are the charge density associated to the original potentials $V(r)$ and the deformed term $V^{\text{def}}(r)$, respectively. Since the problem is strictly two-dimensional we have 
\begin{equation}
\label{ndef}
n^{\text{def}}(\vr)=n^{\text{def}}(r)\delta(z)\,,\qquad
n^{\text{def}}(r)=-\frac{\epsilon\hbar v_F}{e}\frac{1}{r}\frac{d}{dr}\left(r\frac{d}{dr}V^{\text{dif}}(r)\right)\,.
\end{equation}
The distorted electric potential arises from an extra charge distribution on the crystal that can be computed directly from the potential itself. In particular, for the potential  (\ref{paramdef})
one gets 
 \begin{equation}
 n_m^{\text{def}}(\vr)= -\frac{\epsilon \hbar v_F}{e r^3}
 \frac{8 \gamma ^2 \eta^{2 \gamma  } (a_1+a_2\eta^{2\gamma}+a_3\eta^{4\gamma}+a_4\eta^{6\gamma}+a_5\eta^{8\gamma})}
 {\left(
 (2 m+1) \eta^{4 \gamma }-4 \beta  \eta^{2 \gamma }+2 m+1
 \right)^3}\delta(z)\, ,
 \end{equation}
where the constant coefficients are given by 
 \begin{align}
&a_1=(1-2 \gamma )^2 (2 m+1)^2\,,\quad
&a_2= 8\beta(1+2m) (2 \gamma  (\gamma +1)-1)\,,
\\ &a_3= 8 (3\beta ^2 (1-4 \gamma ^2)+\gamma^2(1-12 \gamma ^2)) \,,\quad
&a_4=8 \beta  (2 (\gamma -1) \gamma -1) (2 m+1)\,, \\
&a_5= (2 \gamma +1)^2 (2 m+1)^2\,.
\end{align}

\begin{figure}[H]
\begin{center}
\includegraphics[scale=0.415]{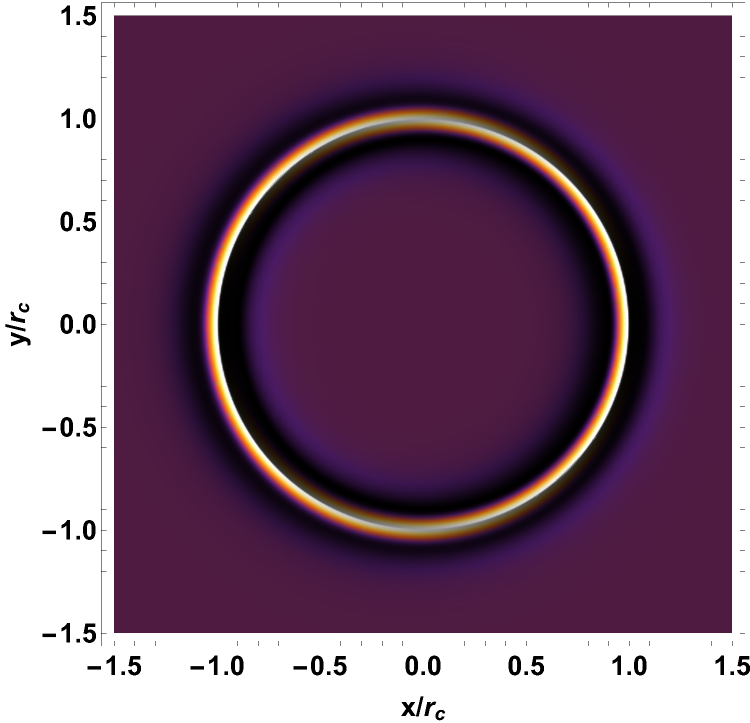}
\includegraphics[scale=0.4]{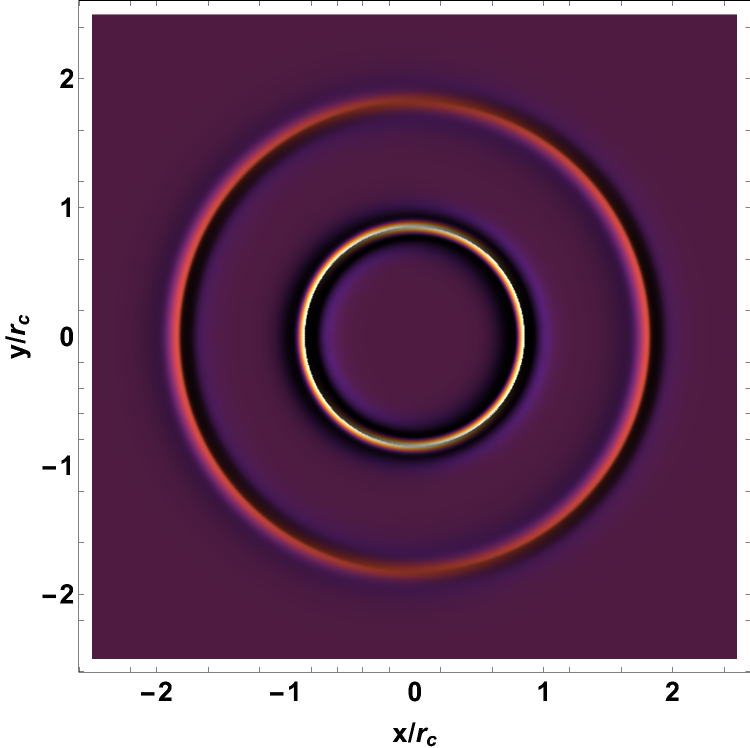}
\includegraphics[scale=0.4]{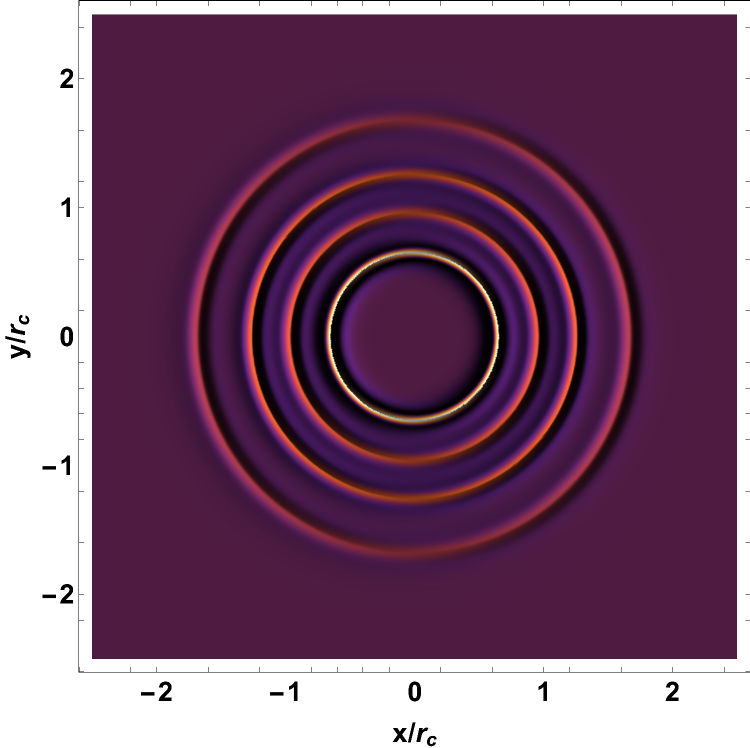}
\end{center}
\caption{\small{
 Plot of the dimensionless charge $\frac{R^3e}{\epsilon \hbar v_F}n^{\text{def}}(r)$ for the potentials constructed in the previous sections.  
}}
\label{XX2}
\end{figure}

The charge density plots corresponding to the potentials from the previous paragraph are shown in figure \ref{XX2}. From them, it is explicitly seen that they correspond to ring-shaped charge accumulations. From here, we interpret the barrier- and well- potentials appearing at each deformation as quantum rings with positive and negative charges, respectively. These rings work as effective traps for charge carriers in the presence of the impurity at the center. From equations (\ref{vmmax}) and (\ref{rinext}), we conclude that the geometry of the rings can be tuned to trap particles in them with a desired angular momentum quantum number. 

\newpage

\section{Discussion and Outlook}
\label{SecDis}
In this article, we presented the method for constructing partially solvable models of Dirac fermions in graphene in the presence of radially symmetric electrostatic fields. Our approach was based on a modification of the standard supersymmetric (Darboux) transformation applied to an intermediate, non-hermitian operator (\ref{alternativePro}), which is finally mapped to a physical Dirac Hamiltonian with a radially symmetric electric field (\ref{Hm}). As a result, the method provides a novel mechanism for constructing the zero energy eigenstates of the new system. In order to test the method developed here, we have created ring-like deformations of the initial potential. Each of these deformations preserves time-reversal symmetry, so they can host a pair of zero-energy bound states per Dirac valley. In addition, the trapped particles are characterized by circular currents flowing in opposite directions.

We focused the analysis on the construction of ring-decorated Coulomb potentials.  The number of deformations in the form of concentric rings and their location in space can be deliberately adjusted by choosing specific combinations of the initial wave functions. This allows to construct analytical potentials under a controlled scenario in the same way as the current experimental setups. Thus, the geometry of the rings is strongly related to the angular momentum quantum number of the particles they can trap, which can be chosen at will. It is worth noting that these deformations are strongly localized, i.e. they do not change the asymptotic behavior of the initial interaction at the origin and at infinity. This means that if we want to introduce controlled deformations of a specific phenomenon, one could just start from a known one. In the particular case of the Coulomb potential, this implies that the atomic collapse states for lower angular momentum states are preserved by the transformation.

Since the presented method can be applied to any initial model, we can take a wide variety of well-known systems as a starting point. A particular example was analyzed in \cite{asym1}, where a connection from the free particle model to the Lorentzian well via the supersymmetric transformation was established, which also shows the shape invariance of such a system. Considering that the freedom is limited only to the initial systems, these results open several possible research directions. It will be interesting to consider the electron-electron interaction and the screening effect associated with the deformations created with our approach in the framework of density functional and Hartree-Fock theories \cite{Kat,singularP2,defunth}. The method can also be used to study the existence of zero energy bound states and their impact on the physics of the system dictated by the Levinson theorem \cite{phaseEQ}, the fermionic cloaking \cite{Cloak1} or the Klein tunneling effect \cite{CorInzJak} in new scenarios. One could also study the effects at the level of probability and current densities when the background geometry is changed, see e.g.\cite{Curved1,Curved2}. Finally, it is possible to look for the generalization of the present method to higher spin equations or to systems without rotational symmetry.

\section*{Acknowledgement}
F.C. and L.I. were supported by Fondecyt Grants No.
1211356 and No. 3220327, respectively. V.J. acknowledges
the assistance provided by the Advanced Multiscale Materials
for Key Enabling Technologies project, supported by the Ministry of Education, Youth, and Sports of the Czech Republic. Project No. CZ.$02.01.01/00/22\_008/0004558$, Co-funded by
the European Union. L. I. also thanks the group ``Urdimbre escritura acad\'emica" for their advice on scientific writing and Dr. Carolina Manquian for enlightening comments regarding the experimental background.

\end{document}